\documentclass[reprint,notitlepage,aps,pra,amsmath,amssymb,showpacs,superscriptaddress]{revtex4-1}
\usepackage{helvet}
\usepackage[T1]{fontenc}
\usepackage[latin9]{inputenc}
\usepackage{mathrsfs}
\usepackage{mathtools}
\usepackage{amsmath}
\usepackage{amssymb}
\usepackage{graphicx}
\usepackage[unicode=true,
 bookmarks=false,
 breaklinks=false,pdfborder={0 0 1},backref=false,colorlinks=false]
 {hyperref}
\hypersetup{pdftitle={Dipole},
 pdfauthor={BM},
 hyperfootnotes=false,linkcolor=black,urlcolor=blue,anchorcolor=black,citecolor=blue,urlcolor=black}
\usepackage{breakurl}

\makeatletter

\usepackage{epstopdf}

\usepackage{amsfonts}\usepackage{latexsym}
\usepackage{amsthm}\usepackage{amsfonts}\usepackage{mathrsfs}\usepackage{bbm}\usepackage{enumerate}
\usepackage{color}
\usepackage{pst-func}
\usepackage{dcolumn}
\usepackage{bm}

\renewcommand{\epsilon}{\varepsilon}

\makeatother

\begin{document}

\title{Leggett mode in a two-component Fermi gas with dipolar interactions}

\author{Brendan C. Mulkerin}

\affiliation{Centre for Quantum and Optical Science, Swinburne University of Technology,
Melbourne 3122, Australia.}

\author{Xia-Ji Liu}

\affiliation{Centre for Quantum and Optical Science, Swinburne University of Technology,
Melbourne 3122, Australia.}

\author{Hui Hu}

\affiliation{Centre for Quantum and Optical Science, Swinburne University of Technology,
Melbourne 3122, Australia.}

\date{\today}
\begin{abstract}
We develop an effective field theory to understand collective modes
of a three-dimensional two-component Fermi superfluid with dipolar
inter-particle interactions, which are modeled by an idealized separable
potential. We first examine the phase transition of the system at
zero temperature, as the fermionic superfluidity is known to be characterized
by two competing order parameters. We find that for strong interactions
there exists a regime where the two order parameters are out-of-phase
and coupled, giving rise to an undamped massive Leggett mode. This
is in addition to the well-known gapless phonon mode. We show that
the Leggett mode can be seen in the spectral function of the in-medium
Cooper pairs, and in principle could be measured through Bragg spectroscopy.
\end{abstract}

\pacs{03.75.Hh, 03.75.Ss, 67.85-d}
\maketitle

\section{Introduction}

Owing to the rapid experimental progress on the control of ultracold
gases over the last decade \cite{Bloch2008,Lahaye2009,Chin2010},
there has been significant work done on creating ultracold dipolar
atomic gases with large magnetic moments \cite{Lu2012,Aikawa2014,Baumann2014,Maier2015,Burdick2016,Baier2018}
and polar molecules with large electronic dipole moments \cite{Ni2008,Ni2010,Wu2012,Tetsu2014,Park2015,Marco2018}.
The long-range and anisotropic nature of the dipole-dipole interaction
in these systems leads to many fascinating quantum phenomena, such
as self-bound droplets in Bose systems \cite{igor2016,Schmitt2016,Chomaz2016},
$p$-wave and topological superfluidity in Fermi systems \cite{Bruun2008,Cooper2009,Levinsen2011},
and quantum chaos \cite{Frisch2014,Baumann2014,Lucie2018}.

There are extensive many-body calculations of Fermi dipolar systems
\cite{you1999,Baranov2002,chan2010,zhao2010}, which mainly focus
on the mean-field regime. To deal with the short-range divergence
of the dipolar interaction, the most common method of using the two-body
$T$-matrix as a way to renormalize the interaction \cite{schmitt1989pairing}
is not tractable in the many-body calculations, since the dipolar
interaction couples different partial wave channels \cite{Yi2000,Kanjilal2008,Bohn2009}.
A useful strategy for renormalization is to take the Born approximation
\cite{Baranov2008,Baranov2012,zhai2013}, which unfortunately is appropriate
in the weakly interacting regime only \cite{Wang2008,Corro2016,Che2016}.
In this work, we consider an effective \emph{separable} interaction
potential that couples different angular momentum channels (i.e.,
$|l-l'|\leq2$) in the strongly interacting regime, as was used in
Ref.~\cite{Shi2013}. Using a separable potential captures the low-energy
physics of the dipolar interaction and allows us to account for the
effect of the coupling between different partial wave channels. In
particular, it provides us a convenient framework to compute the order
parameters for each scattering channel and to explore the behavior
of these order parameters.

For systems with \emph{multiple} superfluid order parameters, there
can exist an additional collective mode other than the well-known
phonon mode, the so-called Leggett mode \cite{Leggett66}. This mode
is characterized as the out-of-phase coupling of different superfluid
order parameters. It has been long predicted to occur in two-band
superconductors, for example in MgB$_{2}$ \cite{Blumberg2007}, and
in non-equilibrium systems \cite{krull2016}. Most recently, an ultracold
atomic Fermi gas near an orbital Feshbach resonances has been thought
to be a possible candidate for exhibiting the Leggett mode \cite{Iskin2016,He2016,Zhang2016}.
The purpose of this work is to show that a two-component Fermi gas
with dipolar interactions provides an excellent new platform to observe
the long-sought Leggett mode.

The rest of the paper is set out as follows. In Sec.~\ref{Sec:many_body}
we consider the many-body thermodynamic potential and derive the mean-field
equations for the density and order parameters. We determine the order
parameters in the different phases of the system as we sweep over
scattering lengths, and examine the symmetry of the associated momentum
distribution. In Sec.~\ref{Sec:calc_deets} we calculate the collective
modes by expanding the thermodynamic potential to second order, which
correspond to the pair fluctuations at the Gaussian level. We show
that the Leggett mode is undamped for a range of interaction strengths
and how the collective modes can be seen in the spectral function
of the Cooper pairs. In Sec.~\ref{Sec.disc} we discuss and summarize
our findings.

\section{Many-body thermodynamic potential}

\label{Sec:many_body}

We consider a many-body two-component Fermi gas with dipolar interactions
in three dimensions, described by the model Hamiltonian (we set $\hbar=1$
and the volume $V$=1) \cite{Gurarie2007,Shi2013}, 
\begin{equation}
\mathcal{H}=\sum_{\mathbf{k}\sigma}\xi_{\mathbf{k}}^{\,}a_{\mathbf{k}\sigma}^{\dagger}a_{\mathbf{k}\sigma}^{\,}+\mathcal{H}_{{\rm int}},
\end{equation}
where the single-particle dispersion is $\xi_{\mathbf{k}}=\mathbf{k}^{2}/2M-\mu$
with the chemical potential $\mu$, $a_{\mathbf{k}\sigma}^{\dagger}\equiv a_{\mathbf{k}\sigma}^{\dagger}(\tau)$
and $a_{\mathbf{k}\sigma}\equiv a_{\mathbf{k}\sigma}(\tau)$ are creation
and annihilation operators respectively, for atoms with spin $\sigma$
and mass $M$, and the interaction Hamiltonian is given by,
\begin{alignat}{1}
\mathcal{H}_{{\rm int}}=\sum_{\mathbf{k}\mathbf{k'}\mathbf{q}}U(\mathbf{k}-\mathbf{k'})a_{\frac{\mathbf{q}}{2}-\mathbf{k}\uparrow}^{\dagger}a_{\frac{\mathbf{q}}{2}+\mathbf{k}\downarrow}^{\dagger}a_{\frac{\mathbf{q}}{2}+\mathbf{k'}\downarrow}^{\;}a_{\frac{\mathbf{q}}{2}-\mathbf{k'}\uparrow}^{\;},
\end{alignat}
where the dipolar interaction is $U(\mathbf{k})=4\pi d^{2}\left(\cos^{2}\theta_{\mathbf{k}}-1\right)/3$,
with $d$ being the dipole moment of the two dipoles polarized along
the $z$-axis and $\theta_{\mathbf{k}}$ the angle between $\mathbf{k}$
and the $z$-axis. We can write the interaction in the following separable
form \cite{Ho2005,Iskin2006prl,iskin2006,Shi2013}, 
\begin{alignat}{1}
U(\mathbf{k}-\mathbf{k'})=4\pi\sum_{j}g_{j}w_{j}(\hat{\mathbf{k}})w_{j}^{*}(\hat{\mathbf{k}}'),\label{eq:dipole_red}
\end{alignat}
where the coupling constants $g_{j}$ satisfy the renormalization
condition for the effective scattering lengths $\lambda_{j}$ \footnote{See Appendix \ref{App:2body} for more details}:
\begin{alignat}{1}
\frac{M}{4\pi\lambda_{j}}=\frac{1}{g_{j}}+\int\frac{d^{3}\mathbf{k}}{(2\pi)^{3}}\frac{M}{\mathbf{k}^{2}}.\label{Eq:renorm}
\end{alignat}
Truncating the sum in Eq.~\eqref{eq:dipole_red} to the two lowest-order
terms as was done in Ref.~\cite{Shi2013}, the effective scattering
lengths are given by 
\begin{alignat}{1}
\lambda_{1,2}=\left[a_{00}\pm{\textrm{sgn}}(a_{02})\sqrt{a_{00}^{2}+4a_{02}^{2}}\right]/2,
\end{alignat}
where $a_{00}$ and $a_{02}$ are the scattering lengths of the $s$
and $d$ partial wave channels. For a set of scattering lengths $(a_{00},a_{02})$,
either $\lambda_{1}$ or $\lambda_{2}$ will be positive, supporting
a bound state energy of $E_{b}=-1/M\lambda_{j}^{2}$~\cite{Shi2013}.
Throughout this work we set $a_{00}^{-1}>0$ and sweep across $a_{02}^{-1}$,
thus there will be a phase transition as the bound state changes
from $\lambda_{1}$ to $\lambda_{2}$ as $a_{02}$ changes sign. The
orthogonal basis vectors in Eq.~\eqref{eq:dipole_red} are given
by \cite{Note1}, 
\begin{alignat}{1}
w_{1,2}(\hat{\mathbf{k}})=\frac{s_{1,2}Y_{00}(\hat{\mathbf{k}})+Y_{20}(\hat{\mathbf{k}})}{\sqrt{s_{1,2}^{2}+1}},
\end{alignat}
where $s_{1,2}=-\left(y\pm\sqrt{y^{2}+4}\right)/2$ and $y=a_{00}/a_{02}$,
and $Y_{lm}(\hat{\mathbf{k}})$ are the spherical harmonics.

The Hamiltonian with the separable potential in Eq.~\eqref{eq:dipole_red}
then becomes,
\begin{alignat}{1}
\mathcal{H}=\sum_{\mathbf{k}\sigma}\xi_{\mathbf{k}}a_{\mathbf{k}\sigma}^{\dagger}a_{\mathbf{k}\sigma}+4\pi\sum_{\mathbf{q},j=(1,2)}g_{j}b_{j}^{\dagger}(\mathbf{q},\tau)b_{j}(\mathbf{q},\tau),
\end{alignat}
where $b_{j}(\mathbf{q},\tau)=\sum_{\mathbf{k}}w_{j}(\hat{\mathbf{k}})a_{-\mathbf{k}+\mathbf{q}/2\uparrow}a_{\mathbf{k}+\mathbf{q}/2\downarrow}$.
In the imaginary time formalism we can write the partition function
as $\mathcal{Z}=\int Da^{\dagger}Da\exp(-S)$, where the action $S$
is given by ($\beta\equiv1/k_{B}T$)
\begin{alignat}{1}
S=\int_{0}^{\beta}d\tau\left[\sum_{\mathbf{k}\sigma}a_{\mathbf{k}\sigma}^{\dagger}(\tau)\partial_{\tau}a_{\mathbf{k}\sigma}(\tau)+\mathcal{H}(\tau)\right].
\end{alignat}
Using the standard Hubbard-Stratonovich transformation, we may decouple
the interaction term by introducing auxiliary complex pairing fields
($j=1,2$), $\Phi_{\mathbf{q}}^{j}(\tau)$. Physically, each pairing
field roughly describes a Cooper pair consisting of two fermions,
i.e., 
\begin{equation}
\Phi_{\mathbf{q}}^{j}(\tau)\sim4\pi g_{j}b_{j}\left(\mathbf{q},\tau\right).
\end{equation}
Using the Nambu spinor representation $\Psi_{\mathbf{k}}^{\dagger}=(a_{\mathbf{k}\uparrow}^{\dagger},a_{-\mathbf{k}\downarrow}^{\,})$
for a two-component Fermi gas, we can rewrite the action as,
\begin{alignat}{1}
S=\int_{0}^{\beta}d\tau\Biggl[ & -\sum_{\mathbf{q},j}\frac{\left|\Phi_{\mathbf{q}}^{j}\left(\tau\right)\right|^{2}}{4\pi g_{j}}\nonumber \\
 & +\frac{1}{2}\sum_{\mathbf{kk}'}\left(\xi_{\mathbf{k}}\delta_{\mathbf{kk}'}-\Psi_{\mathbf{k}}^{\dagger}\mathcal{G}_{\mathbf{kk}'}^{-1}\Psi_{\mathbf{k}'}^{\vphantom{\dagger}}\right)\Biggl],
\end{alignat}
where the inverse fermionic Green's function takes the form ($\mathbf{p}\equiv\frac{\mathbf{k}+\mathbf{k}'}{2}$)
\begin{alignat}{1}
\mathcal{G}_{\mathbf{kk}'}^{-1}=\left[\begin{array}{cc}
-(\partial_{\tau}+\xi_{\mathbf{k}})\delta_{\mathbf{kk}'} & \sum_{j}\Phi_{\mathbf{k}-\mathbf{k}'}^{j}\left(\tau\right)w_{j}\left(\mathbf{\hat{p}}\right)\\
\sum_{j}\Phi_{-\mathbf{k}+\mathbf{k}'}^{j*}\left(\tau\right)w_{j}^{*}\left(\mathbf{\hat{p}}\right) & -(\partial_{\tau}-\xi_{\mathbf{k}})\delta_{\mathbf{kk}'}
\end{array}\right].
\end{alignat}
By integrating out the fermionic degrees of freedom from the partition
function and taking the Fourier transform from imaginary time to Matsubara
frequencies, we obtain the effective action 
\begin{alignat}{1}
S_{{\rm eff}}=-\beta\sum_{Q,j}\frac{|\Phi_{Q}^{j}|^{2}}{4\pi g_{j}}+\sum_{K,K'}\left[\beta\xi_{\mathbf{k}}\delta_{KK'}-{\rm Tr}\ln\mathcal{G}_{KK'}^{-1}\right],\label{eq:Seff}
\end{alignat}
where $Q\equiv(i\nu_{n},\mathbf{q})$ with bosonic Matsubara frequencies
$\nu_{n}=2\pi n/\beta$ and $K\equiv(i\omega_{m},\mathbf{k})$ with
fermionic Matsubara frequencies $\omega_{m}=(2m+1)\pi/\beta$. We
have also used the short-hand notations, $\sum_{Q}\equiv k_{B}T\sum_{i\nu_{n}}\sum_{\mathbf{q}}$
and $\sum_{K}\equiv k_{B}T\sum_{i\omega_{m}}\sum_{\mathbf{k}}$.

In the following, we make a saddle-point approximation and expand
the action in orders of the fluctuation fields $\hat{\phi}_{j}(Q)$
around the order parameters $\Delta_{j}$,
\begin{equation}
\Phi_{Q}^{j}=\Delta_{j}\delta_{Q0}+\hat{\phi}_{j}(Q),
\end{equation}
and we can obtain $S_{{\rm eff}}=S_{{\rm MF}}+S_{{\rm GF}}+\cdots$,
where $S_{{\rm MF}}$ is the mean-field action and $S_{{\rm GF}}$
is the Gaussian fluctuation action.

\subsection{Mean-field theory}

First looking at the mean-field contribution to the action, we have
\begin{alignat}{1}
S_{{\rm MF}}=-\beta\sum_{j}\frac{\left|\Delta_{j}\right|^{2}}{4\pi g_{j}}+\sum_{K}\left[\beta\xi_{\mathbf{k}}-{\rm Tr}\ln\mathcal{G}_{{\rm sp}}^{-1}\right],
\end{alignat}
where the saddle-point Green's function is given by
\begin{alignat}{1}
\mathcal{G}_{\textrm{sp}}^{-1}\left(K\right)=\left[\begin{array}{cc}
i\omega_{m}-\xi_{\mathbf{k}} & \Delta(\mathbf{k})\\
\Delta^{*}(\mathbf{k}) & i\omega_{m}+\xi_{\mathbf{k}}
\end{array}\right],
\end{alignat}
the quasiparticle dispersion is $E_{\mathbf{k}}=\sqrt{\xi_{\mathbf{k}}^{2}+|\Delta(\mathbf{k})|^{2}}$
and we have defined $\Delta(\mathbf{k})=\sum_{j}\Delta_{j}\omega_{j}(\hat{\mathbf{k}})$.
We thus obtain the mean-field thermodynamic potential,
\begin{equation}
\Omega_{{\rm MF}}=-\sum_{j}\frac{\left|\Delta_{j}\right|^{2}}{4\pi g_{j}}+\sum_{\mathbf{k}}\left[\xi_{\mathbf{k}}-E_{\mathbf{k}}-\frac{2}{\beta}\ln\left(1+e^{-\beta E_{\mathbf{k}}}\right)\right],
\end{equation}
and from the condition $\delta\Omega_{{\rm MF}}/\delta\Delta_{j}^{*}=0$
we get the coupled gap equations at finite temperature 
\begin{alignat}{1}
-\frac{\Delta_{j}}{4\pi g_{j}}=\sum_{\mathbf{k},j'}\frac{\Delta_{j'}\omega_{j'}({\hat{\mathbf{k}}})\omega_{j}^{*}({\hat{\mathbf{k}}})}{2E_{\mathbf{k}}}\tanh\frac{\beta E_{\mathbf{k}}}{2}.
\end{alignat}
\begin{figure}
\centering{}\includegraphics[width=0.5\textwidth]{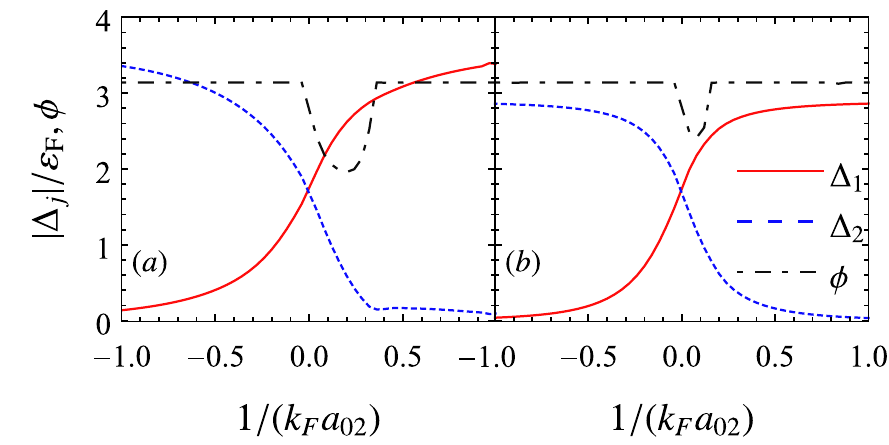}
\caption{Plots of the order parameters $\left|\Delta_{j}\right|$ in units
of the Fermi energy, $\varepsilon_{{\rm F}}$, for a range of $a_{02}$
scattering lengths and (a) $k_{{\rm F}}a_{00}=2$ and (b) $k_{{\rm F}}a_{00}=5$.
The relative phase of the order parameters, $\phi={\rm arg}(\Delta_{1})-{\rm arg}(\Delta_{2})$,
is shown in both plots. \label{fig:Delta}}
\end{figure}
Using the renormalization condition, Eq.~\eqref{Eq:renorm} to replace
the bare coupling constants $g_{j}$, and using the fact that the
basis functions are orthogonal, we rewrite the gap equation at zero
temperature: 
\begin{alignat}{1}
-\frac{M\Delta_{j}}{16\pi^{2}\lambda_{j}}=\sum_{\mathbf{k},j'}\Delta_{j'}\omega_{j'}({\hat{\mathbf{k}}})\omega_{j}^{*}({\hat{\mathbf{k}}})\left(\frac{1}{2E_{\mathbf{k}}}-\frac{M}{\mathbf{k}^{2}}\right).\label{eq:deltas}
\end{alignat}
The number equation at the mean-field level is easily found from the
relation $n=-\partial\Omega_{{\rm MF}}/\partial\mu$: 
\begin{alignat}{1}
n=\int\frac{d^{3}\mathbf{k}}{(2\pi)^{3}}\left(1-\frac{\epsilon_{\mathbf{k}}}{E_{\mathbf{k}}}\right).\label{eq:number}
\end{alignat}
Together the above two equations form a closed set and we can solve
for the chemical potential $\mu$ and order parameters $\Delta_{j}$.
As we take only the first two partial wave channels, the thermodynamic
potential only depends on the absolute values of the order parameters,
$|\Delta_{1}|$ and $|\Delta_{2}|$, and the relative phase between
the two order parameters, $\phi={\rm arg}(\Delta_{1})-{\rm arg}(\Delta_{2})$.
There are several solutions to the number and gap equations for a
given set of scattering lengths, which correspond to different local
minima. The true ground state should be determined by minimizing the
energy density, $\mathcal{E}\equiv\Omega+\mu n$: 
\begin{alignat}{1}
\mathcal{E}(\Delta_{1},\Delta_{2},\mu)=\int\frac{d^{3}\mathbf{k}}{(2\pi)^{3}}\left[\epsilon_{\mathbf{k}}-E_{\mathbf{k}}+\frac{\left|\Delta(\hat{\mathbf{k}})\right|^{2}}{2E_{\mathbf{k}}}\right]+\frac{\mu k_{{\rm F}}^{3}}{3\pi^{2}}.
\end{alignat}

For our units of numerical calculations, we take the Fermi wave vector,
$k_{\textrm{F}}\equiv(3\pi^{2}n)^{1/3}$, as the units of the wave
vectors and the Fermi energy $\varepsilon_{{\rm F}}=\hbar^{2}k_{{\rm F}}^{2}/(2M)$
as the units of energy. This is equivalent to setting $2M=\hbar=1$.
In Fig.~\ref{fig:Delta} we plot the order parameters for a range
of the scattering length $a_{02}$, where we set $k_{{\rm F}}a_{00}=2$
in Fig.~\ref{fig:Delta}(a) and $k_{{\rm F}}a_{00}=5$ in (b). We
see the non-trivial behavior of the order parameters, depending on
the sign of $a_{02}$ and the associated two-body bound state \cite{Shi2013}.
The change of the dominant order parameter implies that the condensate
can have two different symmetries and therefore there exists a quantum
phase transition, as discussed in the previous work \cite{Shi2013}.
Near $(k_{{\rm F}}a_{02})^{-1}\simeq0$, the two order parameters
become comparable. The relative phase of the order parameters also
changes with $k_{{\rm F}}a_{02}$, taking a non-trivial value near
$(k_{{\rm F}}a_{02})^{-1}\simeq0$ for both values of $k_{{\rm F}}a_{00}$.
It is always out-of-phase for $k_{{\rm F}}a_{02}<0$.

\begin{figure}
\centering{}\includegraphics[width=0.5\textwidth]{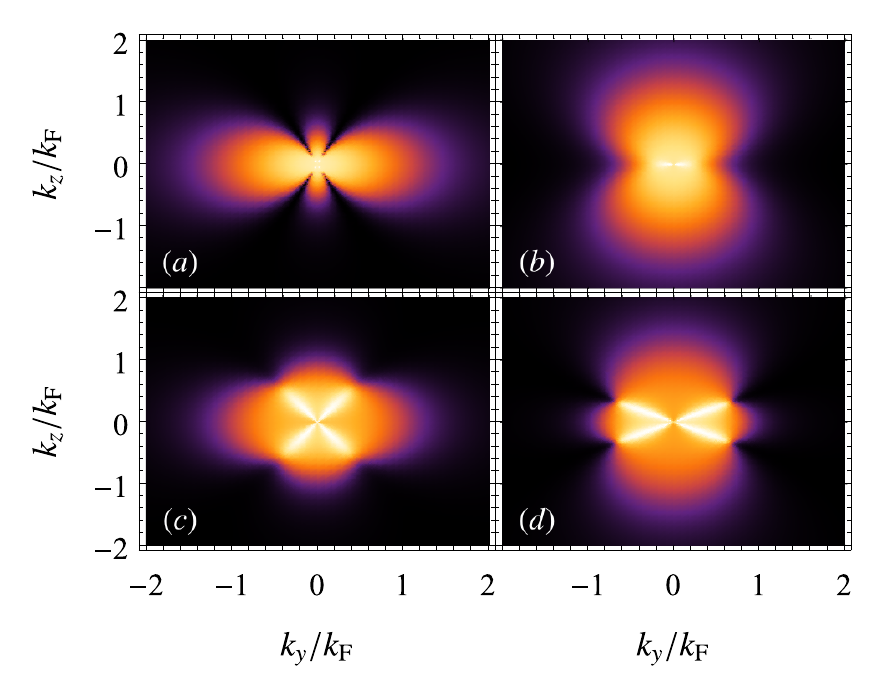}
\caption{Density plots of the momentum distribution $n(k_{x}=0,k_{y},k_{z})=1-\epsilon_{\mathbf{k}}/E_{\mathbf{k}}$
in units of the Fermi momentum, $k_{{\rm F}}$, for different sets
of interaction parameters $(k_{{\rm F}}a_{00},k_{{\rm F}}a_{02})$:
(a) $(1,1)$, (b) $(1,-1)$, (c) $(1,5)$, and (d) $(1,-5)$. \label{fig:density}}
\end{figure}

\subsection{Momentum distribution}

The momentum dependence in the different interaction regimes is non-trivial
due to the mixing of order parameters and angular dependence of the
dipolar interaction. This has already been investigated in the previous
work, by considering the quasiparticle spectral function \cite{Shi2013}.
Here, we show that the momentum distribution can also exhibit different
underlying symmetry, depending on the sign of $k_{{\rm F}}a_{02}$
and the resulting two-body bound state. 

In Fig.~\ref{fig:density} we show the zero-temperature density plots
of the momentum distribution 
\begin{equation}
n\left(k_{x}=0,k_{y},k_{z}\right)=1-\frac{\epsilon_{\mathbf{k}}}{E_{\mathbf{k}}}
\end{equation}
at $k_{{\rm F}}a_{00}=1$ and at different values of $k_{{\rm F}}a_{02}$:
(a) $k_{{\rm F}}a_{02}=1$, (b) $k_{{\rm F}}a_{02}=-1$, (c) $k_{{\rm F}}a_{02}=5$,
and (d) $k_{{\rm F}}a_{02}=-5$. We note that the rotational symmetry
of the system in the $x-y$ plane ensures $n(k_{x}=0,k_{y},k_{z})=n(k_{x},k_{y}=0,k_{z})$.

As the scattering length $k_{{\rm F}}a_{20}$ changes, we see how
the underlying symmetry of the momentum distribution is changing.
In Fig.~\ref{fig:density}(a) the distribution has a $s-d_{z^2}$ like
symmetry and in Fig.~\ref{fig:density}(b) the symmetry is $s+d_{z^2}$. 
For both interactions the momentum distribution is dominated
by the contribution from the bound-state dominated order parameter
in $\Delta(\mathbf{k})$: $\Delta_{1}w_{1}(\hat{\mathbf{k}})$ for
(a) and $\Delta_{2}w_{2}(\hat{\mathbf{k}})$ for (b).

As we increase $|k_{{\rm F}}a_{02}|$, in Figs.~\ref{fig:density}(c)
and (d) we see a higher order non-trivial symmetry. The distribution
is no longer dominated by the bound-state order parameter and the
mixing of order parameters becomes important. For the negative scattering
length, $k_{{\rm F}}a_{02}=-5$, in Fig.~\ref{fig:density}(c) the
coupling of the order parameters is out-of-phase. This is where we
expect the dipolar superfluid to support an additional collective
mode; we will soon see that the system has two collective modes in
this regime. For the positive scattering length, $k_{{\rm F}}a_{02}=5$,
in Fig.~\ref{fig:density}(d) the relative phase of order parameters
becomes non-trivial, and the time reversal symmetry has been broken
due to the order parameter mixing \cite{Shi2013}. In this interaction
regime we expect there to be no Leggett mode as the order parameters
are not out-of-phase.

\section{Collective modes}

\label{Sec:calc_deets}

To study the behavior of the collective modes, we calculate the Gaussian
fluctuation contribution to the effective action, Eq.~\eqref{eq:Seff}.
This can be taken into account by expanding the action to the second
order of the bosonic fields $\hat{\phi}_{j}(Q)$ and $\hat{\phi}_{j}^{*}(Q)$
\cite{Diener2008}: 
\begin{alignat}{1}
S_{{\rm GF}} & =\sum_{Q}\left[-\sum_{j}\frac{\left|\hat{\phi}_{j}\left(Q\right)\right|^{2}}{4\pi g_{j}}\right]+\frac{\beta}{2}\sum_{QK}{\rm Tr}\left[\mathscr{\mathcal{G}}\left(K-\frac{Q}{2}\right)\right.\nonumber \\
 & \left.\times\Phi\left(-Q\right)\mathcal{G}\left(K+\frac{Q}{2}\right)\Phi\left(Q\right)\right],
\end{alignat}
where we have the saddle-point fermionic Green's function and fluctuation
fields, 
\begin{alignat}{1}
\mathcal{G}(K) & =\frac{1}{(i\omega_{m})^{2}-E_{\mathbf{k}}^{2}}\left[\begin{array}{cc}
i\omega_{m}+\xi_{\mathbf{k}} & -\Delta(\hat{\mathbf{k}})\\
-\Delta^{*}(\hat{\mathbf{k}}) & i\omega_{m}-\xi_{\mathbf{k}}
\end{array}\right],\\
\Phi(Q) & =\left[\begin{array}{cc}
0 & \sum_{j}\hat{\phi}_{j}(-Q)\omega_{j}(\hat{\mathbf{k}})\\
\sum_{j}\hat{\phi}_{j}^{*}(Q)\omega_{j}^{*}(\hat{\mathbf{k}}) & 0
\end{array}\right].
\end{alignat}
The subscript ``sp'' in the saddle-point Green's function has been
suppressed for a better presentation. From this we can then rearrange
the terms to obtain the final form, 
\begin{alignat}{1}
S_{{\rm GF}}=\frac{\beta}{2}\sum_{Q,jj'}\left[\hat{\phi_{j}}^{*}(Q),\hat{\phi}_{j}^{\,}(-Q)\right]M_{jj'}(Q)\left[\begin{array}{c}
\hat{\phi}_{j'}^{\,}(Q)\\
\hat{\phi}_{j'}^{*}(-Q)
\end{array}\right],
\end{alignat}
where we have defined the elements $M_{jj'}$ (each of which is a
2 by 2 matrix), 
\begin{widetext}
\begin{alignat}{1}
\left[M_{jj'}\right]{}_{11}(Q) & =\sum_{K}\mathcal{G}_{11}\left(\frac{Q}{2}+K\right)\mathcal{G}_{22}\left(\frac{Q}{2}-K\right)\omega_{j}^{\vphantom{*}}(\hat{\mathbf{k}})\omega_{j'}^{*}(\hat{\mathbf{k}})-\frac{\delta_{j,j'}}{4\pi g_{j}},\\
\left[M_{jj'}\right]{}_{12}(Q) & =\sum_{K}\mathcal{G}_{12}\left(\frac{Q}{2}+K\right)\mathcal{G}_{12}\left(\frac{Q}{2}-K\right)\omega_{j}^{\vphantom{*}}(\hat{\mathbf{k}})\omega_{j'}^{*}(\hat{\mathbf{k}}),
\end{alignat}
$[M_{jj'}]_{21}(Q)=[M_{jj'}]_{12}(Q)$, and $[M_{jj'}]_{22}(Q)=[M_{jj'}]_{11}(-Q)$.
We then complete the sums over the Matsubara frequencies to arrive
at the zero-temperature result: 
\begin{alignat}{1}
\left[M_{jj'}\right]{}_{11}(Q) & =-\frac{\delta_{jj'}}{4\pi g_{j}}+\sum_{\mathbf{k}}\left(\frac{u_{-}^{2}u_{+}^{2}}{i\nu_{n}-E_{+}-E_{-}}-\frac{v_{+}^{2}v_{-}^{2}}{i\nu_{n}+E_{+}+E_{-}}\right)\omega_{j}^{\vphantom{*}}(\hat{\mathbf{k}})\omega_{j'}^{*}(\hat{\mathbf{k}}),\\
\left[M_{jj'}\right]{}_{12}(Q) & =-\sum_{\mathbf{k}}\left(\frac{u_{+}v_{+}u_{-}v_{-}}{i\nu_{n}-E_{+}-E_{-}}-\frac{u_{+}v_{+}u_{-}v_{-}}{i\nu_{n}+E_{+}+E_{-}}\right)\omega_{j}^{\vphantom{*}}(\hat{\mathbf{k}})\omega_{j'}^{*}(\hat{\mathbf{k}}),
\end{alignat}
where we define the BCS parameters $u_{\pm}^{2}=\left(1+\xi_{\pm}/E_{\pm}\right)/2$
and $v_{\pm}^{2}=\left(1-\xi_{\pm}/E_{\pm}\right)/2$, and the short-hand
notations $\xi_{\pm}=\xi_{\mathbf{k}\pm\mathbf{q}/2}$ and $E_{\pm}=E_{\mathbf{k}\pm\mathbf{q}/2}$.
We renormalize the bare coupling constants $g_{j}$ again using Eq.~\eqref{Eq:renorm}
and this also cures the divergences in the integrals of $M_{11}$.
We can then write the \emph{inverse} boson propagator for Cooper pairs,
$\mathbf{M}(Q)$, as a $2N_{j}\times2N_{j}$ matrix, where $N_{j}$
is the number of channels and in this work $N_{j}=2$. 
\end{widetext}

Analytically continuing the Matsubara frequencies, $i\nu_{n}\rightarrow\omega+i0^{+}$,
the phonon and Leggett collective mode dispersions are determined
by the equation ${\rm det}\left[\mathbf{M}(\mathbf{q},\omega)\right]=0$.
As the scattering potential we have used for the dipole-dipole interaction
has an angular dependency, the bosonic propagator $\mathbf{\Gamma}(\mathbf{q},\omega)=\mathbf{M}^{-1}(\mathbf{q},\omega)$
has an angular dependence and is a function of three parameters: $\mathbf{\Gamma}(\mathbf{q},\omega)\equiv\mathbf{\Gamma}(q,\theta,\omega)$,
where $q\equiv\left|\mathbf{q}\right|$ and $\theta\equiv\theta_{\mathbf{q}}$.

\begin{figure}
\centering{}\includegraphics[width=0.5\textwidth]{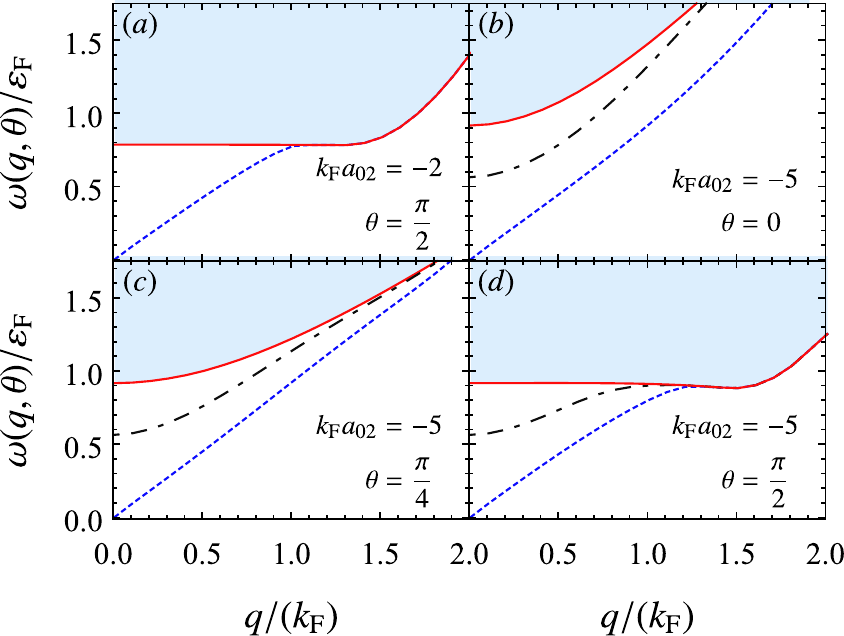}
\caption{Plots of the two-particle continuum (blue shaded region), phonon mode
(blue dashed), and Leggett mode (black dash-dotted) in units of the
Fermi energy for scattering lengths $k_{{\rm F}}a_{00}=2$ and (a)
$(k_{{\rm F}}a_{02},\theta)=(-2,\pi/2)$, (b) $(k_{{\rm F}}a_{02},\theta)=(-5,0)$,
(c) $(k_{{\rm F}}a_{02},\theta)=(-5,\pi/4)$, and (d) $(k_{{\rm F}}a_{02},\theta)=(-5,\pi/2)$.
\label{fig:modes}}
\end{figure}

\subsection{Results}

We plot in Fig.~\ref{fig:modes} the collective modes for scattering
lengths $k_{{\rm F}}a_{00}=2$ and (a) $(k_{{\rm F}}a_{02},\theta)=(-2,\pi/2)$,
(b) $(k_{{\rm F}}a_{02},\theta)=(-5,0)$, (c) $(k_{{\rm F}}a_{02},\theta)=(-5,\pi/4)$,
and (d) $(k_{{\rm F}}a_{02},\theta)=(-5,\pi/2)$. The two-particle
continuum is shown as the blue shaded region, the phonon mode is the
blue-dashed line, and the Leggett mode is the black-dot-dashed line.

We see in Fig.~\ref{fig:modes}(a) that the dipolar superfluid supports
only the phonon mode when the channel coupling $a_{02}$ is weak,
which becomes damped once it enters the two-particle continuum as
we increase the momentum $q/k_{{\rm F}}$, indicating that for this
interaction regime the system is BCS like \cite{tempere_d_wave,Shi2013}.
Looking at Fig.~\ref{fig:Delta}(a) for scattering lengths $(k_{{\rm F}}a_{00},k_{{\rm F}}a_{02})=(2,-2)$,
the order parameters are out-of-phase and the superfluid is mainly
characterized by the order parameter $\Delta_{2}$, thus we would
expect the mixing between the order parameters to be negligible and
there is no Leggett mode.

In Figs.~\ref{fig:modes}(b)-\ref{fig:modes}(d) we increase the
channel coupling to $k_{{\rm F}}a_{02}=-5$, and we now see two undamped
collective modes, the Leggett and phonon modes, at low momentum. In
this interaction regime the order parameters are approximately at
the same order of magnitude and are out-of-phase (see Fig. \ref{fig:Delta}(a),
where $1/(k_{{\rm F}}a_{02})=-0.2$), satisfying Leggett's original
picture of two well-defined and coupled condensates \cite{Leggett66}.
The Leggett mode merges into the two-particle continuum and becomes
damped for large momentum at each $\theta$. The phonon mode is always
undamped for $\theta=0$, but it merges into the two-particle continuum
for large momenta when $\theta$ becomes sufficiently large; see,
for example, Fig.~\ref{fig:modes}(d).

\begin{figure}
\centering{}\includegraphics[width=0.5\textwidth]{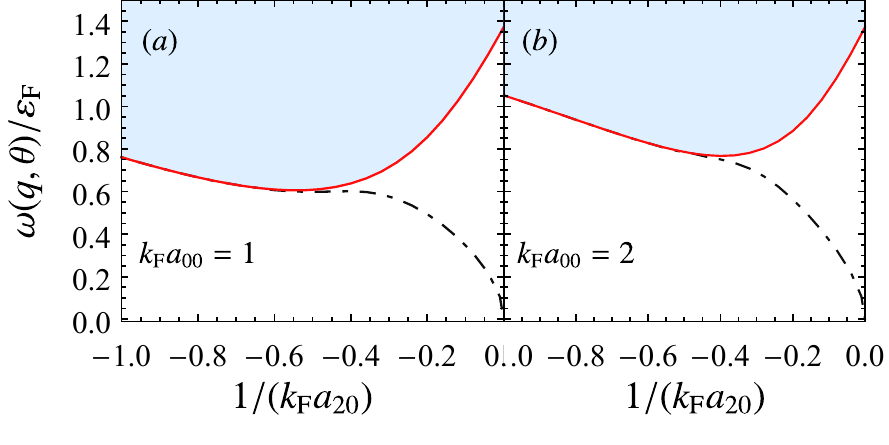}
\caption{Plots of the two particle continuum (blue shaded region) and Leggett
mode (black dash-dotted) for $q=0$ as a function of the scattering
length $k_{{\rm F}}a_{02}$ at (a) $k_{{\rm F}}a_{00}=1$ and (b)
$k_{{\rm F}}a_{00}=2$. \label{fig:legmodes}}
\end{figure}

In Figs.~\ref{fig:legmodes}(a) and \ref{fig:legmodes}(b) we plot
only the Leggett mode (black dash-dotted line) and the two-particle
continuum (blue shaded region) for a range of $k_{{\rm F}}a_{02}$
at zero momentum $q=0$, and set $k_{{\rm F}}a_{00}=1$ and $k_{{\rm F}}a_{00}=2$,
respectively. For negative scattering length, $(k_{{\rm F}}a_{02})^{-1}<0$,
we see in both figures the Leggett mode becomes undamped for large
enough $|k_{{\rm F}}a_{02}|$, and disappears as the scattering length
changes sign. Here, the system undergoes a quantum phase transition
as the bound state changes its character and the relative phase between
the two order parameters starts to deviate from $\pi$. We find for
positive scattering lengths, $(k_{{\rm F}}a_{02})^{-1}>0$, there
are no longer two collective modes and the Leggett mode always lies
in the two-particle continuum (not shown in the figure). This can
be understood from Fig.~\ref{fig:Delta}: for large positive $k_{{\rm F}}a_{02}$,
the relative phase of the order parameters exhibits non-trivial dependence
on $(k_{\textrm{F}}a_{02})^{-1}$ and is not completely out-of-phase.
As $(k_{{\rm F}}a_{02})^{-1}$ increases further, the two order parameters
become out-of-phase again, however $\Delta_{2}$ becomes dominant
and leaves no room for the Leggett mode.

\begin{figure}
\centering{}\includegraphics[width=0.5\textwidth]{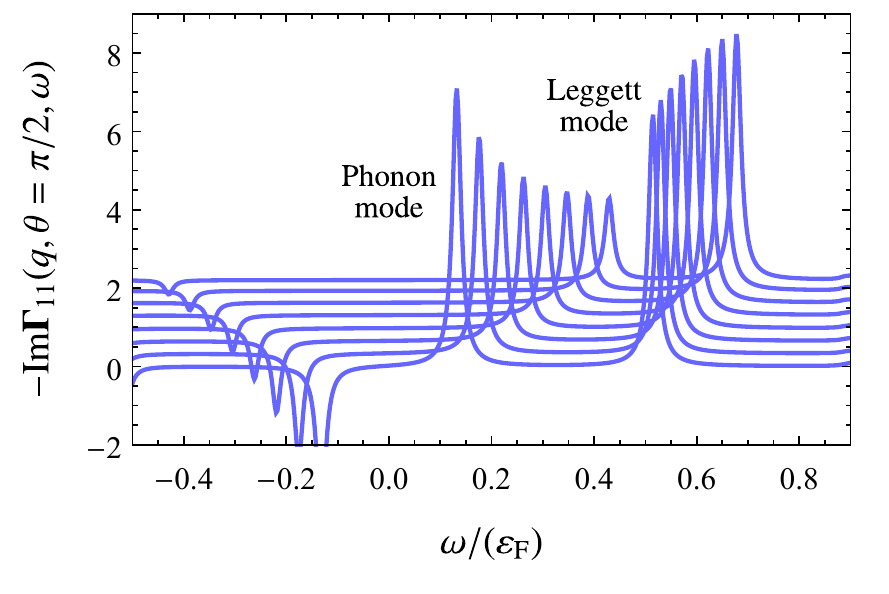}
\caption{ Plot of the spectral function of Cooper pairs, $-{\rm Im}\mathbf{\Gamma}_{11}(q,\theta,\omega)$,
in arbitrary units for momenta $q=0.1k_{{\rm F}}$ to $0.5k_{{\rm F}}$,
scattering lengths $(k_{{\rm F}}a_{00},k_{{\rm F}}a_{02})=(1,-5)$,
and $\theta=\pi/2$.\label{fig:imgam}}
\end{figure}

Experimentally, the collective modes of a strongly interacting Fermi
gas can be probed by measuring the density dynamic structure factor
via Bragg spectroscopy \cite{Combescot2006,Veeravalli2008}. We would
expect that, if the regimes where the Leggett mode is undamped can
be reached, we should be able to measure the phonon  and Leggett
modes. To support this idea, in Fig.~\ref{fig:imgam} we show a typical
spectral function of the in-medium Cooper pairs, i.e., $-{\rm Im}\mathbf{\Gamma}_{11}(q,\theta,\omega)$,
in arbitrary units for a range of momenta, where for clarify we have
shifted each curve to be visible. We have chosen an interaction strength
of $k_{{\rm F}}a_{00}=1$ and $k_{{\rm F}}a_{02}=-5$. We can clearly
see how the phonon mode and Leggett mode evolve as the momentum increases.

\section{Discussion and summary}

\label{Sec.disc}

We have found that an undamped Leggett mode requires interactions
in the $k_{{\rm F}}a_{00}$ and $k_{{\rm F}}a_{02}$ channels to be
such that both order parameters are significant and out-of-phase.
Practically, such an interaction regime could be achieved with a multichannel
resonance \cite{Kanjilal2008,Shi2013,Rafa2016}, changing the scattering
lengths by sweeping across the shape resonances induced by the dipolar
interaction. For polar molecule systems, the large electronic dipole
moments can be adjusted such that the interaction regime to observe
the Leggett mode could be reached \cite{Ni2010,Marco2018}. For atomic
species with a magnetic dipole moment, the interaction is fixed but
Feshbach resonances can be used to tune the background $s$-wave interaction,
i.e., the scattering length $a_{00}$. However, as we require a large $k_{{\rm F}}a_{02}$
as well to have a significant coupling between the two channels, a
direct observation of the Leggett mode would be difficult. The addition
of higher order channels would not significantly alter our results,
since the higher order channel coupling will most likely be weak \cite{Kanjilal2008}.

In summary, through an effective separable form of the dipolar interaction
we have investigated the collective modes of a dipolar Fermi gas,
in which the superfluid is described by two order parameters. We have
found for strong interactions, where the order parameters are strongly
coupled and out-of-phase, an additional collective mode - the Leggett
mode - emerges, on top of the phonon mode. We have determined the
interaction regime where this mode persists and have shown that the
Leggett mode can be seen through the spectral function of the Cooper
pairs, indicating that in principle it could be measured through Bragg
spectroscopy.
\begin{acknowledgments}
We thank Jia Wang and Jing Zhou for their comments. Our research was
supported by Australian Research Council's (ARC) Discovery Projects:
DP140100637, FT140100003 and DP180102018 (XJL), FT130100815 and DP170104008
(HH). 
\end{acknowledgments}

\appendix

\section{Two-body scattering}

\label{App:2body}

To renormalize the many-body equations we need to calculate the two-body
T-matrix. The dipolar interaction is non-separable and this makes
the many-body calculation intractable. We separate the dipolar interaction
using the effective potential in Ref.~\cite{Shi2013}, which was
introduced to model a multichannel resonance, and here we briefly
derive the effective potential. The scattering amplitude for the dipolar
interaction is given by \cite{Bohn2009,Yi2000} 
\begin{alignat}{1}
\left.f(\mathbf{k}',\mathbf{k})\right|_{k=k'}=4\pi\sum_{lml'm'} & i^{l'-l}k^{-1}\left(\frac{1}{\mathcal{K}^{-1}-i}\right)_{lm}^{l'm'}\nonumber \\
 & Y_{lm}(\hat{\mathbf{k}})Y_{l'm'}^{*}(\hat{\mathbf{k}}'),
\end{alignat}
where $\mathcal{K}_{lm}^{l'm}$ is the K-matrix and can be calculated
as in Ref.~\cite{Kanjilal2008,Deb2001}. The K-matrix is related
to the T-matrix by $\mathcal{T}=2\left(\mathcal{K}^{-1}-i\right)^{-1}$
and in the small $k$ limit the scattering lengths are given by the
K-matrix elements, $a_{ll'}^{(m)}=-\lim_{k\rightarrow0}\mathcal{K}_{lm}^{l'm}/k$.
Introducing a matrix $\mathcal{A}$ whose elements are defined by
the scattering lengths as $\mathcal{A}_{ll'}^{(m)}=i^{l-l'}a_{ll'}^{(m)}$,
we diagonalize the matrix $\mathcal{A}$ in an orthonormal basis,
$w_{jm}(\hat{\mathbf{k}})=\sum_{l}d_{jl}Y_{lm}(\hat{\mathbf{k}})$
\footnote{Since we will set $m=0$ these eigenfunctions are real}.
We can write the scattering amplitude as \footnote{We can bring the $i^{l'-l}$ term down into the fraction as it will
not affect the identity matrix.}, 
\begin{alignat}{1}
\left.f(\mathbf{k}',\mathbf{k})\right|_{k=k'\rightarrow0}=4\pi\sum_{jm}f_{jm}w_{jm}(\hat{\mathbf{k}})w_{jm}^{*}(\hat{\mathbf{k}}'),\label{Eq:Scatt_amp_eff1}
\end{alignat}
where $f_{jm}=-1/(\lambda_{jm}^{-1}+ik)$. We can find a separable
potential which reproduces this scattering amplitude as 
\begin{alignat}{1}
U(\hat{\mathbf{k}}',\hat{\mathbf{k}})=4\pi\sum_{jm}g_{jm}w_{jm}(\hat{\mathbf{k}})w_{jm}^{*}(\hat{\mathbf{k}}'),
\end{alignat}
where the coupling constants $g_{jm}$ satisfies the renormalization
condition, 
\begin{alignat}{1}
\frac{M}{4\pi\lambda_{jm}}=\frac{1}{g_{jm}}+\int\frac{d^{3}\mathbf{k}}{(2\pi)^{3}}\frac{M}{\mathbf{k}^{2}}.\label{Eq:renorm_app}
\end{alignat}
Taking the separable potential to second order as the minimal model
to describe the dipolar interaction, we set the scattering matrix
to the following form for the multichannel resonance, 
\begin{alignat}{1}
A_{sc}=\left(\begin{array}{cc}
a_{00} & -a_{02}\\
-a_{02} & 0
\end{array}\right).\label{eq:Asc}
\end{alignat}
The eigenvalues of this matrix are given by 
\begin{alignat}{1}
\lambda_{1,2}=\left[a_{00}\pm{\textrm{sgn}}(a_{02})\sqrt{a_{00}^{2}+4a_{02}^{2}}\right]/2,
\end{alignat}
and for any set of values $(a_{00},a_{02})$ either $\lambda_{1}$
or $\lambda_{2}$ will be positive with a bound state energy of $E_{b}=-1/M\lambda_{i}^{2}$.
This will mean that as we sweep across $a_{00}^{-1}$ or $a_{02}^{-1}$
there will be a phase transition, since the bound state changes from
$\lambda_{1}$ to $\lambda_{2}$. The choice of Eq.~\eqref{eq:Asc}
is not unique and we can change the sign of the off diagonal elements,
this would have the effect of changing the sign of the $\lambda_{1}$
and $\lambda_{2}$ and would not qualitatively change any of the results
here. The orthogonal basis vectors are given by, 
\begin{alignat}{1}
w_{1,2}(\hat{\mathbf{k}})=\frac{s_{1,2}Y_{00}(\hat{\mathbf{k}})+Y_{20}(\hat{\mathbf{k}})}{\sqrt{s_{1,2}^{2}+1}},
\end{alignat}
with $s_{1,2}=-\left(y\pm\sqrt{y^{2}+4}\right)/2$ and $y=a_{00}/a_{02}$.

\bibliographystyle{apsrev4-1}
\bibliography{dipole}

\end{document}